\documentclass[prd,showpacs,preprintnumbers,amsmath,amssymb,10pt]{revtex4}
\usepackage{graphics}
\usepackage[utf8]{inputenc}
\usepackage{epsfig}
\usepackage{subfigure}
\usepackage{dcolumn}% Align table columns on decimal point
\usepackage{bm}% bold math
\usepackage{color}

\begin{document}

\title{Hunting for the $X_b$ via  hidden bottomonium decays}

\author{Gang Li,$^{1,2}$\footnote{gli@mail.qfnu.edu.cn} and Zhu Zhou$^1$\footnote{zhouzhumail@163.com}
}

\affiliation{$^1$Department of Physics, Qufu Normal University, Qufu 273165, China\\
$^2$State Key Laboratory of Theoretical Physics, Institute of Theoretical Physics, Chinese Academy of Sciences, Beijing 100190, China}

\begin{abstract}
In this work, we study the isospin conserved hidden bottomonium decay of $X_b\to \Upsilon(1S)\omega$, where $X_b$ is taken to be the counterpart of the famous $X(3872)$ in the bottomonium sector as a candidate for the meson-meson molecule. Since it is likely that  the $X_b$ is below the $B\bar B^*$ threshold and the mass difference between the neutral and charged  bottom meson is small compared to the binding energy of the $X_b$, the isospin violating decay mode $X_b\to \Upsilon (nS)\pi^+\pi^-$ would be  greatly suppressed. We use the effective Lagrangian based on the heavy quark symmetry to explore the rescattering mechanism of $X_b\to \Upsilon(1S)\omega$ and  calculate the partial widths. Our results show that the partial width for the $X_b\to \Upsilon(1S)\omega$ is about tens of keVs. Taking into account the fact that the total width of $X_b$ may be smaller than a few MeV like $X(3872)$, the calculated branching ratios may reach to orders of $10^{-2}$. These hidden bottomonium decay modes are of great importance in the experimental search for the $X_b$ particularly  at the hadron collider. Also, the associated studies of hidden bottomonium decays $X_b \to \Upsilon(nS) \gamma$, $\Upsilon(nS)\omega$, and $B\bar B \gamma$ may help us investigate the structure of $X_b$ deeply. The experimental observation of $X_b$ will provide us with further insight into the spectroscopy of exotic states and is helpful to probe  the structure of the states connected  by the heavy quark symmetry.

\end{abstract}

\date{\today}


%14.40.Rt Exotic mesons

%13.75.Lb Meson-meson interactions

%13.20.Gd Decays of J/\psi, and other quarkonia

%14.40.Pq Heavy quarkonia

%14.40.Lb Charmed mesons

\maketitle

\section{Introduction}
\label{sec:introduction}

During recent years, the experimental observation of a large number of so-called $XYZ$ states has inspired tremendous
effort to unravel their nature beyond the conventional quark model~\cite{Brambilla:2010cs,Swanson:2006st,Eichten:2007qx,Voloshin:2007dx,Godfrey:2008nc,Drenska:2010kg,Bodwin:2013nua}.
In $2003$, the Belle Collaboration first reported the observation of $X(3872)$  in the $J/\psi \pi^+\pi^-$ invariant mass spectrum  of $B^+\to K^++ J/\psi \pi^+\pi^-$~\cite{Choi:2003ue} which was subsequently  confirmed
by the BaBar Collaboration~\cite{Aubert:2004ns} in the same channel.
It was also discovered in proton-proton/antiproton
collisions by the D0~\cite{Abazov:2004kp}, CDF~\cite{Aaltonen:2009vj}, and
LHCb Collaborations~\cite{Chatrchyan:2013cld,Aaij:2013zoa}.  The $X(3872)$ is the first, and perhaps the most renowned exotic candidate; however, its nature is still ambiguous due to its peculiar properties. First, only an upper bound of the total width has been measured experimentally:
$\Gamma<1.2$ MeV~\cite{Agashe:2014kda}, which is tiny compared to typical hadronic widths. Second, the mass is located close to the
$D^0\overline D^{*0}$ threshold,
$M_{X(3872)}-M_{D^0}-M_{D^{*0} }=(-0.12\pm0.24)$~MeV~\cite{TheBABAR:2013dja}, which leads  to
speculation that  the $X(3872)$  is presumably  a meson-meson molecular state~\cite{Tornqvist:2004qy,Hanhart:2007yq}.

Many studies on the production and decay of  the $X(3872)$ have been carried out in order to understand its nature. The discrimination of a compact multiquark configuration and a loosely bound hadronic molecule configuration is one important aspect.
From this point,  it is also valuable  to look for the counterpart in the bottom sector, denoted as $X_b$  following the notation in Ref.~\cite{Hou:2006it}, as states related by heavy quark symmetry may have  universal behaviors.
Since the $X_b$ is very heavy and its quantum numbers $J^{PC}=1^{++}$, it is less likely discovered at the current electron-positron collision facilities, though the Super KEKB may provide an opportunity in $\Upsilon(5S,6S)$ radiative decays~\cite{Aushev:2010bq}.

The production of $X_b$ at the LHC and the Tevatron~\cite{Guo:2014sca} and other exotic states at hadron colliders~\cite{Bignamini:2009sk,Artoisenet:2009wk,
Artoisenet:2010uu,Esposito:2013ada,Ali:2011qi,Ali:2013xba,Guo:2013ufa} have been extensively investigated.
On one hand, it is shown that the production rates at the LHC and the Tevatron are sizeable~\cite{Guo:2014sca}. On the other hand, the search for $X_b$ also depends on reconstructing the $X_b$, which motivates us to study the  $X_b$ decays. In Ref.~\cite{Li:2014uia}, we have studied the radiative decays of $X_b\to \gamma \Upsilon(nS)$ ($n=1,2,3$), with $X_b$ being a candidate for the meson-meson molecular state, and found that the partial widths into $\gamma \Upsilon(nS)$ are about $1$ keV. Since this meson is expected to be far below threshold and the mass difference between neutral and charged $B$ mesons is very small, the isospin-violating decay mode, for instance $X_b\to \Upsilon\pi^+\pi^-$, is highly suppressed, and this may explain the escape of $X_b$ in the recent CMS search~\cite{Chatrchyan:2013mea}. As a consequence, the isospin-conserved decays $X_b\to \Upsilon(1S) \omega$, on which we will focus this paper, will be of high priority.

To calculate $X_b\to \Upsilon(1S) \omega$, we investigate the intermediate meson-loop (IML) contributions.
As is well known, IML transitions have been one of the important nonperturbative
transition mechanisms in many processes, and their impact on the heavy quarkonium transitions has
been noticed for a long time~\cite{Lipkin:1986bi,Lipkin:1986av,Lipkin:1988tg,Moxhay:1988ri}. Recently, this mechanism has been applied to study $B$ decays~\cite{Cheng:2004ru,Lu:2005mx}, the production and decays of exotic states~\cite{Wang:2013cya,Li:2014gxa,Bondar:2011ev,Cleven:2013sq,Chen:2011zv,Chen:2013coa,Li:2012as,Chen:2011pv,Li:2013yla,Zhao:2013jza,Guo:2010ak,Guo:2009wr,Liu:2013vfa,Guo:2013zbw,
Wang:2013hga,Voloshin:2013ez,Voloshin:2011qa,Guo:2010zk,Chen:2011pu,Chen:2012yr,Chen:2013bha} and a global agreement with experimental data were obtained. Thus, this may be an effective approach to deal with the $X_b$ hidden bottomonium decays.

The rest of this paper is organized as follows. In Sec.~\ref{sec:formula}, we introduce the formalism used in this work. Numerical results are presented in Sec.~\ref{sec:results}, and the summary is given in
Sec.~\ref{sec:summary}.

%%%%%%%%%%%%%%%%%%%%%
\section{Hidden bottomonium decays}
\label{sec:formula}
%%%%%%%%%%%%%%%%%%%%%

%%%%%%%%%%%%%%%%%%%%%%%%%%%%%%
%%%%%%%%%%%%%%%%%%%%%%%%%%%%%%
\begin{figure}[hbt]
\begin{center}
\includegraphics[width=0.7\textwidth]{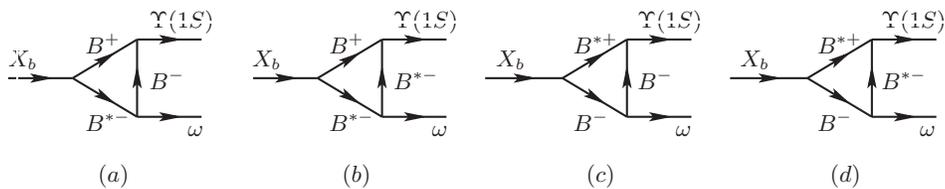}
\caption{Feynman  diagrams for $X_b \to \Upsilon(1S) \omega$ with  the $B{\bar B}^*$ as the intermediate states.} \label{fig:loops}
\end{center}
\end{figure}
%%%%%%%%%%%%%%%%%%%%%%%%%%%%%%
A state of the heavy meson pair can be decomposed in terms of the total spin of the $Q\bar Q$ pair $S_H$ and the total angular momentum $S_L$ of the rest degrees of freedom, for which the latter includes both the total spin of the light quark pair $q\bar q$ and any orbital momentum~\cite{Voloshin:2011qa,Bondar:2011ev,Li:2013kya,Li:2013yka,Wang:2014wga,Voloshin:2004mh,Voloshin:2012dk}. In the S-wave $B{\bar B}^*$ molecular picture of $X_b$, the heavy quark spin structure is mandated to be $1^{--}_{H}\otimes 1^{--}_{L}$~\cite{Voloshin:2004mh}. However, in the tetraquark picture of $X_b$, the total spin $S_H$ can be either $0$ or $1$.
As a result, the dominant hidden bottomonium decay channel is $\Upsilon(1S) \omega$ in the molecule picture of $X_b$, while the hidden bottomonium decay channels of $X_b$ can be those channels with $S_H=0$ or $1$ final bottomonium.  From this point of view, it seems that a molecule $X_b$ may have a bigger branching ratio of $X_b \to \Upsilon(1S) \omega$ than a tetraquark $X_b$.

Generally speaking, we should include all the possible intermediate meson loops in the calculation. In
reality, the breakdown of the local quark-hadron duality allows us to pick up the leading contributions as a reasonable approximation~\cite{Lipkin:1986bi,Lipkin:1986av}. Since we consider the isoscalar $X_b$ as an $S$-wave molecular state with $J^{PC} = 1^{++}$ given by the superposition of $B^0 {\bar B}^{*0}+c.c.$ and $B^- {B}^{*+}+c.c.$ hadronic configurations as
\begin{eqnarray}
|X_b\rangle= \frac {1} {2} [  (|B^0{\bar B}^{*0}\rangle - |B^{*0} {\bar B}^0\rangle) +   (| B^+ B^{*-}\rangle - | B^- B^{*+}\rangle ) ],
\end{eqnarray}
we will only consider the $BB^*$ meson loops here. The coupling of $X_b$ to the bottomed meson pair is based on the following effective  Lagrangian,
\begin{eqnarray}
{\cal L} = \frac {1} {2} X_{b \mu}^{\dagger} [x_1(B^{*0 \mu} {\bar B}^0 - B^{0} {\bar B}^{*0 \mu})+x_2(B^{*+\mu} B^- - B^+ B^{*-\mu})] + H.c.,
\end{eqnarray}
where $x_i$ denotes the coupling constant. For the $X_b$  below the $S$-wave $BB^*$ threshold, the effective coupling of this state is related to the probability of finding the two-hadron component in the physical wave function of the bound states and the binding energy, $E_{X_b}=m_B+m_{B^*}-m_{X_b}$~\cite{Weinberg:1965zz, Baru:2003qq,Guo:2013zbw}
\begin{eqnarray}\label{eq:coupling-Xb}
x_i^2 \equiv 16\pi (m_B+ m_{B^*})^2 c_i^2 \sqrt{\frac {2E_{X_b}}{\mu}} ,
\end{eqnarray}
where $c_i=1/{\sqrt 2}$, $\mu=m_Bm_{B^*}/(m_B+m_{B^*})$ is the reduced mass. Here, the coupling constant $x_i$ in Eq.~(\ref{eq:coupling-Xb}) is based on the assumption that $X_b$ is a shallow bound state where the potential binding the mesons is mostly contact range.

Based on the heavy quark symmetry, the leading-order effective Lagrangian for the $\Upsilon(1S)$ can be expressed as~\cite{Colangelo:2003sa,Casalbuoni:1996pg}
\begin{eqnarray}
%%Upsilon B(*)B(*)
\mathcal{L}_{\Upsilon(1S) B^{(*)} B^{(*)}} &=&
ig_{\Upsilon BB} \Upsilon_{\mu} (\partial^\mu B \bar{B}- B
\partial^\mu \bar{B})-g_{\Upsilon B^* B} \varepsilon_{\mu \nu
\alpha \beta}
\partial^{\mu} \Upsilon^{\nu} (\partial^{\alpha} B^{*\beta} \bar{B}
 + B \partial^{\alpha}
\bar{B}^{*\beta})\nonumber\\
&&-ig_{\Upsilon B^* B^*} \big\{
\Upsilon^\mu (\partial_{\mu} B^{* \nu} \bar{B}^*_{\nu}
-B^{* \nu} \partial_{\mu}
\bar{B}^*_{\nu})+ (\partial_{\mu} \Upsilon_{\nu} B^{* \nu} -\Upsilon_{\nu}
\partial_{\mu} B^{* \nu}) \bar{B}^{* \mu}  \nonumber\\
&& +
B^{* \mu}(\Upsilon^\nu \partial_{\mu} \bar{B}^*_{\nu} -
\partial_{\mu} \Upsilon^\nu \bar{B}^*_{\nu})\big\}, \label{eq:h1}
\end{eqnarray}
where
${{B}^{(*)}}=\left(B^{(*)+},B^{(*)0}\right)$ and
${\bar B^{(*)T}}=\left(B^{(*)-},\bar{B}^{(*)0}\right)$ correspond to the
bottom meson isodoublets. $\epsilon_{\mu\nu\alpha\beta}$ is the antisymmetric Levi-Civita tensor and $\epsilon_{0123}= +1$. Due to the heavy quark symmetry, the following relationships of the couplings can be used~\cite{Casalbuoni:1996pg,Colangelo:2003sa}
\begin{eqnarray}
g_{\Upsilon(1S) BB} = 2g_1 \sqrt{m_{\Upsilon(1S)}} m_B \ ,
\quad g_{\Upsilon(1S) B^* B} = \frac {g_{\Upsilon(1S) BB}} {\sqrt{m_B m_{B^*}}} \ ,
\quad g_{\Upsilon(1S) B^* B^*} = g_{\Upsilon(1S) B^* B}  \sqrt{\frac {m_{B^*}} {m_B}} m_{B^*},
\end{eqnarray}
where $g_1 = \sqrt{m_{\Upsilon(1S)}}/(2m_B f_{\Upsilon(1S)})$,
$m_{\Upsilon(1S)}$ and $f_{\Upsilon(1S)}$ denote the mass and decay constant of
$\Upsilon(1S)$, respectively. The decay constant $f_{\Upsilon(1S)}$ can
be extracted from the $\Upsilon(1S)\to e^+e^-$,
\begin{eqnarray}
\Gamma(\Upsilon(1S) \to e^+e^-) = \frac {4\pi\alpha_{em}^2} {27} \frac {f_{\Upsilon(1S)}^2} {m_{\Upsilon(1S)}},
\end{eqnarray}
where $\alpha_{em} = 1/137$ is the electromagnetic fine-structure constant. Using the mass and
leptonic decay width of  the $\Upsilon(1S)$ state, $\Gamma(\Upsilon(1S)
\to e^+e^-) =1.340 \pm 0.018$ keV~\cite{Agashe:2014kda}, one can  obtain $f_{\Upsilon(1S)} =
715.2 $ {\rm MeV}.

For the Lagrangians relevant to the light vector mesons used in this work, we can write them in the chiral and heavy quark limits~\cite{Casalbuoni:1996pg,Cheng:2004ru},
\begin{eqnarray}
{\cal L} &=& - ig_{\mathcal{B}\mathcal{B}\mathcal{V}} \mathcal{B}_i^\dagger {\stackrel{\leftrightarrow}{\partial}}{\!_\mu} \mathcal{B}^j(\mathcal{V}^\mu)^i_j -2f_{\mathcal{B}^*\mathcal{B}\mathcal{V}} \epsilon_{\mu\nu\alpha\beta}
(\partial^\mu \mathcal{V}^\nu)^i_j
(\mathcal{B}_i^\dagger{\stackrel{\leftrightarrow}{\partial}}{\!^\alpha} \mathcal{B}^{*j \beta} -\mathcal{B}_{i}^{*\beta \dagger}{\stackrel{\leftrightarrow}{\partial}}{\!^\alpha} \mathcal{B}^j) + ig_{\mathcal{B}^*\mathcal{B}^*\mathcal{V}} \mathcal{B}^{*\nu\dagger}_i {\stackrel{\leftrightarrow}{\partial}}{\!_\mu} \mathcal{B}^{*j}_\nu(\mathcal{V}^\mu)^i_j\nonumber \\
&&
+4if_{\mathcal{B}^*\mathcal{B}^*\mathcal{V}} \mathcal{B}^{*\dagger}_{i\mu}(\partial^\mu \mathcal{V}^\nu-\partial^\nu
\mathcal{V}^\mu)^i_j \mathcal{B}^{*j}_\nu  \, , \label{eq:LDDV}
\end{eqnarray}
where the heavy meson couplings to light vector mesons have the following
relationships,
\begin{eqnarray}
g_{\mathcal{B}\mathcal{B}{\cal V}}=g_{\mathcal{B}^*\mathcal{B}^*{\cal V}}=\frac{\beta
g_V}{\sqrt2}, \quad
f_{\mathcal{B}^*\mathcal{B}{\cal V}}&=&\frac{f_{\mathcal{B}^*\mathcal{B}^*{\cal V}}}{m_{\mathcal{B}^*}}=\frac{\lambda g_V}{\sqrt2}\, ,
\end{eqnarray}
with the parameters $\beta=0.9$, $\lambda = 0.56 \,
{\rm GeV}^{-1}$, and $g_V = {m_\rho /
f_\pi}$~\cite{Casalbuoni:1996pg,Isola:2003fh,Becirevic:2009yb}.

Based on the relevant Lagrangians given above, the loop transition amplitudes in
Fig.~\ref{fig:loops} can be expressed
in a general form as follows,
\begin{eqnarray}
{\cal A}_{fi}=\int \frac {d^4 q_2} {(2\pi)^4} \sum_{B^* \ \mbox{pol.}}
\frac {T_1T_2T_3} {a_1 a_2 a_3}{\cal F}(m_2,q_2^2)\, ,
\end{eqnarray}
where $T_i$ and $a_i = q_i^2-m_i^2 \ (i=1,2,3)$ are the vertex
functions and the denominators of the intermediate meson
propagators, respectively. For example, in Fig.~\ref{fig:loops} (b), $T_i \
(i=1,2,3)$ are the vertex functions for the initial $X_b$, final
bottomonium and final $\omega$, respectively. $a_i \
(i=1,2,3)$  are the denominators for the intermediate $B^+$,
$B^{*+}$ and $B^{* -}$ propagators, respectively.
In order to take care of the off-shell effects of the exchanged particles~\cite{Li:1996yn,Locher:1993cc,Li:1996cj}, we adopt a dipole form factor
\begin{eqnarray}\label{ELA-form-factor}
{\cal F}(m_{2}, q_2^2) \equiv \left(\frac
{\Lambda^2-m_{2}^2} {\Lambda^2-q_2^2} \right)^2,
\end{eqnarray}
with $\Lambda\equiv m_2+\alpha\Lambda_{\rm QCD}$, and the QCD energy
scale $\Lambda_{\rm QCD} = 220$ MeV.  Many phenomenological studies have suggested the cutoff parameter $\alpha\sim 2$. The explicit expression of transition for $X_b \to \Upsilon(1S) \omega$ amplitudes is given in Appendix~\ref{appendix-A}.
\section{Numerical Results}
\label{sec:results}
%%%%%%%%%%%%%%%%%%%%%%%
\begin{table}[htb]
\begin{center}
\caption{Predicted partial widths (in units of keV) of the $X_b$ decays. The parameter in the form factor is chosen as $\alpha =2.0$, $2.5$, and $3.0$, respectively.} \label{tab:results-1}
\begin{tabular}{cccc}
\hline
Dipole form factor & $\alpha=2.0$  & $\alpha=2.5$ & $\alpha=3.0$  \\ \hline
$E_{X_b}=1$ MeV  & $4.03$  & $8.55$  & $15.53$ \\ \hline
$E_{X_b}=5$ MeV  & $8.38$  & $17.84$ & $32.51$ \\ \hline
$E_{X_b}=10$ MeV & $11.17$  & $23.84$ & $43.56$  \\ \hline
$E_{X_b}=25$ MeV & $15.12$  & $33.30$ & $61.10$  \\ \hline
$E_{X_b}=50$ MeV & $18.63$  & $40.14$ & $73.96$  \\ \hline
$E_{X_b}=100$ MeV & $20.02$  & $43.34$ & $80.22$  \\ \hline
\end{tabular}
\end{center}
\end{table}
%%%%%%%%%%%%%%%%%%%%%%%

%%%%%%%%%%%%%%%%%%%%%%%%%%%%%%
%%%%%%%%%%%%%%%%%%%%%%%%%%%%%%
\begin{figure}[hbt]
\begin{center}
\includegraphics[width=0.7\textwidth]{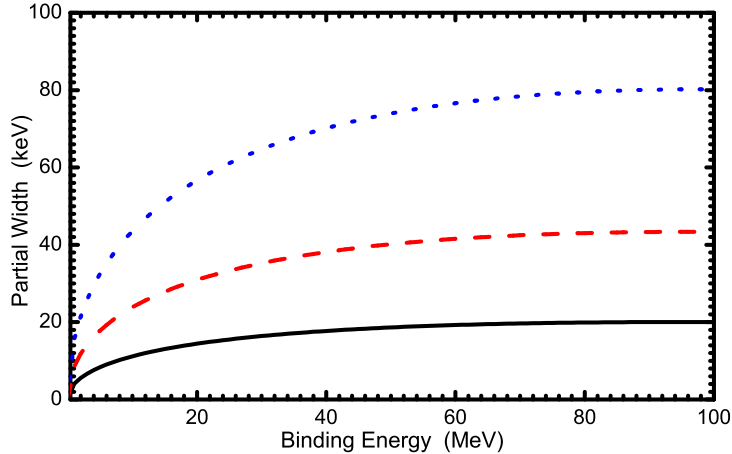}
\caption{The dependence of partial widths of $X_b\to  \Upsilon(1S) \omega$ on the $E_{X_b}$ using dipole form factors with $\alpha =2.0$ (solid lines), $\alpha=2.5$ (dashed lines), and $\alpha=3.0$ (dotted lines), respectively.} \label{fig:WidthOnEXb}
\end{center}
\end{figure}
%%%%%%%%%%%%%%%%%%%%%%%%%%%%%%
Both the tetraquark model~\cite{Ali:2009pi} and hadronic molecular
calculations~\cite{Tornqvist:1993ng,Guo:2013sya,Karliner:2013dqa} have predicted the existence of the $X_b$. In Ref.~\cite{Ali:2009pi}, A. Ali {\it et al.} predicted the mass of the lowest-lying $1^{++}$ $\bar b \bar q bq$ tetraquark to be $10504$ MeV, while the mass of the $B\bar B^*$
molecule is a few tens of MeV
higher~\cite{Guo:2013sya,Karliner:2013dqa}. In Ref.~\cite{Guo:2013sya}, the
mass was predicted to be $(10580^{+9}_{-8})$~MeV, which corresponds to a binding
energy $(24^{+8}_{-9})$~MeV. These studies can provide a range for the binding energy, and in the following we will choose a few illustrative values: $E_{X_b} =(1, 5, 10, 25, 50, 100)$ MeV.

In Table~\ref{tab:results-1}, we list the predicted partial widths by choosing the dipole form factor and three values for the cutoff parameter $\alpha$. From this table, we can see that the widths for the $X_b \to \Upsilon(1S) \omega$ are about tens of keVs. The Particle Data Group gives an upper bound for the total width of $X(3872)$ to be $1.2$ MeV~\cite{Agashe:2014kda}. If we consider the similarity between $X_b$ and $X(3872)$, the full width for $X_b$ should be narrow. Hence, our results would indicate a sizeable branching fraction for $X_b \to \Upsilon(1S) \omega$.

In Fig.~\ref{fig:WidthOnEXb}, we present the partial widths for the  $X_b\to \Upsilon(1S)\omega$ in terms of the  $E_{X_b}$ with the dipole form factors  $\alpha=2.0$ (solid lines), $2.5$ (dashed line), and $3.0$ (dotted lines), respectively. As shown in this figure, there is no cusp structure in the curve. This is because the mass
of $X_b$ lies below the intermediate $BB^*$ threshold. The uncertainties caused by the form factors indicate  our limited knowledge on the applicability of the effective Lagrangian.
However, fortunately, the dependence of the partial widths are not drastically sensitive, which indicates a reasonable cutoff of the ultraviolet contributions by the empirical form factors.
Both the coupling strength of $X_b$ in Eq.~(\ref{eq:coupling-Xb}) and the threshold effects influence the binding energy dependence of the partial width. On the one hand, as shown in Eq.~(\ref{eq:coupling-Xb}), the coupling strength of $X_b$ increases with the binding energy $E_{X_b}$. On the other hand, the threshold effects decrease with $E_{X_b}$.  Both the coupling strength of $X_b$ and the threshold effects change quickly in the small $E_{X_b}$ region and slowly in the large $E_{X_b}$ region. As a result, the behavior of the partial width is relatively sensitive at small $E_{X_b}$, while it becomes smooth at large $E_{X_b}$.

As mentioned above, the coupling constant $x_i$ in Eq.~(\ref{eq:coupling-Xb}) is based on the assumption of a shallow bound state, so our determination of this coupling has uncertainties which are mainly due to the nonperturbative pion exchange. For the interaction between two heavy hadrons forming a bound state, one can formulate an effective field theory (EFT) description~\cite{Guo:2013sya} of hadronic molecules, which is analogous to that of the EFT formulation of the nucleon-nucleon interaction~\cite{Epelbaum:2008ga}. There are two main simplifications for heavy hadrons: The first one is that heavy quark symmetry severely constrains the low-energy interactions among them~\cite{Mehen:2011yh,AlFiky:2005jd}; The second is that pion exchanges are usually perturbative and only produce small effects~\cite{Fleming:2007rp,Valderrama:2012jv}. The only exception is the isoscalar in the bottom sector where the pions might be nonperturbative due to the large masses of the bottom mesons~\cite{Guo:2013sya,Mehen:2011yh,Valderrama:2012jv}.  In Ref.~\cite{Guo:2013sya},  the studies indicate that the relative errors of $C_{0X}$ are about 20\% for the isoscalar $X_b$.  So even if we consider the impact of this effect, the estimated order of the magnitude for the branching ratio $X_b\to \Upsilon(1S) \omega$ may also be sizeable and can be measured in experiments.

In the heavy quark symmetry limit, one may correlate the transition matrix element of the $X_b\to \Upsilon(1S)\omega$ with that of $X(3872)\to J/\psi \omega$.  Generally speaking, with the experimental measurements of the masses of $X(3872)$ and the experimental lower limit branching ratio of $X(3872) \to J/\psi \omega$~\cite{Agashe:2014kda}, one can first use the heavy quark symmetry to determine the cutoff parameter value $\alpha$ and then give a round estimation of the possible experimental measurements on $X_b \to \Upsilon(1S) \omega$. However,  we should also notice that the phase space dependence behavior of the partial widths may be quite different for these two decay channels. For the process $X(3872) \to J/\psi \omega$, the allowed phase space is very small, while the phase space is large enough for the process $X_b \to \Upsilon(1S) \omega$. Maybe systematic experimental and theoretical studies on these decay processes can test this point.

In the case of hidden bottomonium decays of $X_b$ (such as $X_b \to \Upsilon(1S) \omega$ and $\Upsilon(nS) \gamma$) with $X_b$ being an $S$-wave $B {\bar B}^*$ molecule candidate, since the heavy quarks in the $B {\bar B}^*$ meson pair have to recombine to form the final bottomonium, the transition from the $B {\bar B}^*$ meson pair into the $\Upsilon$ plus $\omega$ (or photon) occurs at a distance much smaller than both the size of the $X_b$ as a hadronic molecule and the range of forces between the $B$ and $\bar B^*$ mesons. As a consequence, these processes are not sensitive to the $B {\bar B}^*$ wave function at long distances which is governed by the binding energy, but rather they are determined by the short distance part of the $X_b$. As shown in Ref.~\cite{Li:2014uia}, the radiative decays $X_b\to \Upsilon(nS) \gamma$, and especially their ratio $R$, are not sensitive to the long-range structure of the $X_b$, and thus they cannot be used to rule out the picture that the $X_b$ is dominantly a hadronic molecule. In order to further investigate the structure of $X_b$, it is necessary to explore other decay channels, and the radiative decay $X_b \to B\bar B \gamma$ might give the best hint about the molecular nature of $X_b$. In the case of $X_b \to B\bar B \gamma$, one of the constituent hadrons ($B$) is in the final state and the rest of the final particles are products of the decay of the other constituent hadron (${\bar B}^*$) of the $X_b$ molecule. In these processes, the relative distance between the $B\bar B^*$ pair can be as large as allowed by the size of the $X_b$, since the final state is produced by the decay of ${\bar B}^*$ instead of the rescattering transition. So this process is sensitive to the long-distance part of the wave function of a hadronic molecule and can be used to probe the long structure of $X_b$. The experimental observation of $X_b$, hunting for hidden bottomonium decays ($X_b \to \Upsilon(nS) \gamma$, $\Upsilon(nS) 3\pi$) and $X_b \to B\bar B \gamma$ can help us understand the structure of $X_b$.

\section{Summary}
\label{sec:summary}
%%%%%%%%%%%%%%%%%%%%%%%

In this work, we have investigated the isospin-conserved hidden bottomonium decay $X_b\to \Upsilon(1S)\omega$, where $X_b$ is taken to be the counterpart of $X(3872)$ in the bottomonium sector as a meson-meson molecule candidate. Since the mass of this state may be far below the $B\bar B^*$ threshold, the isospin-violating decay channel $X_b\to \Upsilon\pi^+\pi^-$ would be highly suppressed and stimulate the importance of the isospin-conserved decay channel $X_b\to \Upsilon(1S)\omega$.  We explore the rescattering mechanism with the effective Lagrangian based on the heavy quark symmetry. The calculated partial width of $X_b\to \Upsilon(1S)\omega$ is about tens of keVs. Taking into account the fact that the total width of $X_b$ may be smaller than a few MeV like $X(3872)$, the calculated branching ratios may reach to orders of $10^{-2}$.
This study of hidden bottomonium decay along with the previous work on production rates in hadron-hadron collisions~\cite{Guo:2014sca} and radiative decays of $X_b$~\cite{Li:2014uia} have indicated  a promising  prospect to find the $X_b$ at hadron colliders, in particular, the LHC, and we suggest that our experimental colleagues perform an analysis. Also, the associated studies of hidden bottomonium decays $X_b \to \Upsilon(nS) \gamma$, $\Upsilon(nS)\omega$, and $B\bar B \gamma$ may help us investigate the structure of $X_b$ systematically. Such attempt  will likely lead to the discovery of the $X_b$ and enrich the exotics spectrum in the heavy quarkonium sector.

%%%%%%%%%%%%%%%%%%
\section*{Acknowledgements}
\label{sec:acknowledgements}
The authors thank Xiao-Hai Liu, Qian Wang, and Wei Wang for useful discussions. This work is supported by the National Natural Science Foundation of China (Grant No. 11275113).

%%%%%%%%%%%%%%%%%%

\begin{appendix}

\label{appendix-A}
\section{The Transition Amplitude in ELA}

In this appendix, we give the transition amplitudes for the
intermediate heavy meson loops in Fig.~\ref{fig:loops} in the framework of the ELA. $p_1$, $p_2$, and $p_3$ are the four-vector momenta for initial state $X_b$, final heavy quarkonium $\Upsilon(1S)$, and final light vector meson $\omega$, respectively. $q_1$, $q_2$ and $q_3$ are the four-vector momenta for the intermediate heavy mesons. $\varepsilon_1$, $\varepsilon_2$ and $\varepsilon_3$ are the polarization vectors for initial state, final quarkonium, and final light vector meson, respectively.

\begin{eqnarray}
{\cal A}_{BB^* [B]}&=& \int \frac {d^4q_2} {(2\pi)^4}[g_{X_{b}}
\varepsilon_{1\mu} ] [g_{\Upsilon(1S) BB}(q_1-q_2)_\rho \varepsilon_2^{*\rho}] [-2f_{B^*BV} \varepsilon_{\lambda \theta \phi\kappa} p_3^\lambda \varepsilon_3^{* \theta} (q_2+q_3)^\phi]
\nonumber \\
&& \times \frac {i} {q_1^2-m_1^2}  \frac {i} {q_2^2-m_2^2}  \frac {i(-g^{\mu\kappa}
+q_3^\mu q_3^\kappa/m_3^2)} {q_3^2-m_3^2} {\cal F}(m_{2}, q_2^2), \nonumber \\
{\cal A}_{BB^* [B^*]} &=& \int \frac {d^4q_2} {(2\pi)^4}[g_{X_{b}}
\varepsilon_{1\mu}] [ -g_{\Upsilon(1S) B^*B}\varepsilon_{\rho \sigma \xi\tau}p_2^\rho \varepsilon_2^{*\sigma}q_2^\xi] [2g_{B^*B^*V} q_{2\theta} \varepsilon_3^{* \theta} g_{\phi\kappa} + 4f_{B^*B^*V} (p_{3}^{\theta}\varepsilon_{3\phi}^* - p_{3\phi}\varepsilon_{3}^{*\theta})g_{\kappa \theta}] \nonumber \\
&& \times \frac {i} {q_1^2-m_1^2}  \frac {i(-g^{\tau\phi}
+q_2^\tau q_2^\phi/m_2^2)} {q_2^2-m_2^2}  \frac {i(-g^{\mu\kappa}
+q_3^\mu q_3^\kappa/m_3^2)} {q_3^2-m_3^2} {\cal F}(m_{2}, q_2^2), \nonumber \\
{\cal A}_{B^*B [B]} &=& \int \frac {d^4q_2} {(2\pi)^4}[g_{X_{b}}
\varepsilon_{1\mu}] [ -g_{\Upsilon(1S) B^*B}\varepsilon_{\rho \sigma \xi\tau}p_2^\rho \varepsilon_2^{*\sigma}q_1^\xi ] [-g_{BBV} (q_2+q_3)_\lambda \varepsilon_3^{*\lambda}] \nonumber \\
&& \times \frac {i(-g^{\mu\tau} +q_1^\mu q_1^\tau/m_1^2)}
{q_1^2-m_1^2}  \frac {i} {q_2^2-m_2^2}  \frac {i} {q_3^2-m_3^2} {\cal F}(m_{2}, q_2^2) \nonumber \\
{\cal A}_{B^*B [B^*]}&=& \int \frac {d^4q_2} {(2\pi)^4}[g_{X_{b}}
\varepsilon_{1\mu} ] [g_{\Upsilon(1S) B^*B^*}\varepsilon_2^{*\rho} ((q_2-q_1)_\rho g_{\sigma\xi}+p_{2\xi} g_{\rho\sigma}+p_{2\sigma} g_{\rho\xi})][2f_{B^*BV} \varepsilon_{\lambda \theta \phi\kappa} p_3^\lambda \varepsilon_3^{* \theta} (q_2+q_3)^\phi]
 \nonumber \\
&& \times \frac {i(-g^{\mu\sigma}
+q_1^\mu q_1^\sigma/m_1^2)} {q_1^2-m_1^2}  \frac {i(-g^{\xi\kappa}
+q_2^\xi q_2^\kappa/m_2^2)} {q_2^2-m_2^2}  \frac {i} {q_3^2-m_3^2} {\cal F}(m_{2}, q_2^2), \nonumber \\
\end{eqnarray}

\end{appendix}

\end{document}